\documentclass[twocolumn]{article}

\usepackage[T1]{fontenc}
\usepackage[utf8]{inputenc}

\usepackage{graphicx}

\usepackage{hyperref}

\usepackage{listings}
\AtBeginDocument{}

\usepackage{amsmath,amssymb}
\usepackage{multirow}
\usepackage{subfig}

\begin{document}

\title{Time-limited Bloom Filter
\footnote{This work is partially financed by National Funds through the Portuguese funding agency, FCT - Fundação para a Ciência e a Tecnologia, within project UIDB/50014/2020. This version extends the 4 page version published in ACM SAC 2023 and adds a section on Experimental Evaluation.}}

\author{Ana Rodrigues \and Ariel Shtul \and Carlos Baquero \and Paulo Sérgio Almeida}
\date{June 9, 2023}





\maketitle

\begin{abstract}
A \emph{Bloom Filter} is a probabilistic data structure designed to check, rapidly and memory-efficiently, whether an element is present in a set. It has been vastly used in various computing areas and several variants, allowing deletions, dynamic sets and working with sliding windows, have surfaced over the years.

When summarizing data streams, it becomes relevant to identify the more recent elements in the stream. However, most of the sliding window schemes consider the most recent items of a data stream without considering time as a factor. While this allows, e.g., storing the most recent 10000 elements, it does not easily translate into storing elements received in the last 60 seconds, unless the insertion rate is stable and known in advance. 

In this paper, we present the \emph{Time-limited Bloom Filter}, a new BF-based approach that can save information of a given time period and correctly identify it as present when queried, while also being able to retire data when it becomes stale. The approach supports variable insertion rates while striving to keep a target false positive rate. 
We also make available a reference implementation of the data structure as a Redis module.
\end{abstract}





%
%
%
%

\section{Introduction}

Nowadays, there are several settings where searches of small amounts of information are made in large pools of data stored somewhere. Often, there is an aim to optimize that search, making it a low latency and high throughput operation, by trying to find new data structures, technologies and mechanisms.
    
A \emph{Bloom Filter (BF)} is a hash coding method with allowable errors that can be used for ``testing a series of messages one-by-one for membership in a given set of messages''~\cite{bf}. In more recent years, the BF scheme has been receiving a lot of attention, with many variants surfacing, and is now being used in a wide range of systems/applications, such as web caches~\cite{webcaches-cbf}, networking~\cite{networking3,networking1,networking2} and databases~\cite{databases}

Many of the BF-based approaches consider the most recent elements of the data stream, i.e., a specified number of fresh items is stored. However, none of them take time into account. This is important for many real world scenarios, e.g., to avoid showing the user the same commercial advertisement more than once in a given time period or to be able to check the IPs that connected to a system at a certain time, as well as for fraud detection and prevention of denial of service attacks.

An \emph{Age-Partitioned Bloom Filter (APBF)}~\cite{apbf} is a BF-based data structure able to hold a specified window of elements and evict those that are older. 
In this paper, we present the \emph{Time-limited Bloom Filter}, a variant of the APBF method that forgets information at a given time-based rate, but still, according to the setup of the filter, provides high accuracy when querying a specific time window (e.g., the last minute or the last hour).


\section{Bloom Filters}
\label{sec:bf}

A \emph{Bloom Filter}~\cite{bf} is a space-efficient data structure designed to represent a set of elements and check for membership on that set. In its simple form, a BF consists of a bit array of size \emph{m}, with each bit initially set to $0$. When an element is inserted, \emph{k} bits of the array are set to $1$ by a set of \emph{k} uniform independent hash functions ($h_1, h_2, \ldots, h_k$). To query for its presence, all bits to which the item is hashed to are checked. If at least one bit is $0$, then we are certain the element is not in the BF, otherwise, if all bits are $1$, one considers the element to be present with a certain error probability, known as \emph{false positive rate}.

Usually the memory footprint of BF is defined according to the number of elements to store and to the allowed false positive rate.


\section{Data streams and window models}
\label{sec:window_models}

Since the size of a data stream may be infinite, it is essential to have a mechanism that helps to control which part of the stream is important to the problem at hand. Many BF solutions revisit the concept of window models to process these streams, the most common ones being the \emph{Landmark Window}~\cite{stbf,windows}, \emph{Sliding Window}~\cite{windows,dcba} and \emph{Jumping Window}~\cite{windows,jw}.

A Landmark Window handles disjoint segments of the data stream, one at a time, each limited by a specific landmark (a time interval, e.g., an hour or a day). Since it only stores a segment of the entire data stream at a time, it requires less space than the other two models. However, it fails to establish element relationships between windows, i.e., two duplicates can be missed, if one of them occurs at the end of a landmark and the other at the beginning of the next. 

A Sliding Window considers only the most recent \emph{N} elements of the data stream, which means that for every new element arriving, an old one must be evicted. It is ideal for studying the data stream behavior in real time. For this scheme, any data structure can be used as long as it allows the deletion of elements.

The basic idea of the Jumping Window is to slide the window in jumps as the data flows, by breaking the stream into smaller disjoint sub-windows of fixed size. It ensures the freshness of the results and does not need to store the whole window. However, it cannot accurately represent the data stream actual state, since the number of elements varies as the window jumps. This scheme requires the use of data structures that can combine and subtract efficiently their results.

\section{Duplicate detection in streams}
\label{sec:duplicate_detection}

Duplicate detection is an important operation in many real world scenarios, such as in URL crawling~\cite{urlcrawling1,urlcrawling2}, to avoid the constant fetching of the same URL, and in click streams~\cite{windows}, for fraud detection. Nowadays, there are many different approaches to detect duplicates in a data stream, which are essentially based on \emph{Bloom Filters} or \emph{Dictionaries}.

\subsection{BF-based}

Most approaches for detecting duplicates in a stream of elements are BF-based and basically consist of mapping \emph{k} cells to update using \emph{k} hash functions. Loosely, they can be based on \emph{counters}, \emph{segments} or \emph{timestamps}.

Counter-based methods were originally introduced with the purpose of allowing the deletion of data in a set. The \emph{Stable Bloom Filter}~\cite{stbf} is an example of a counter-based approach that ages the filter by randomly choosing some counters (if greater than $0$) to decrement at every insertion. However, this scheme introduces false negatives and doesn't guarantee that the elements being expired are actually the oldest in the filter.

Segmented-based approaches make use of more than one segment. \emph{Double Buffering}~\cite{dbuff} uses two buffers, an \emph{active} and a \emph{warm-up}, where the first holds the more recent data and the other is a subset of the first. When the \emph{active} becomes full, the two buffers switch roles and the now \emph{warm-up} buffer is cleared out to receive fresher elements. Somewhat similar, \emph{$A^2$ Buffering}~\cite{a2buff} also uses two buffers, \emph{active1} and \emph{active2}, but simultaneously. The first buffer stores the more recent data and the second holds older recent elements. When the \emph{active1} becomes full, everything in \emph{active2} is cleared out and the two buffers switch roles. Comparing these two schemes, $A^2$ Buffering is more memory efficient, since both buffers store distinct elements, while Double Buffering introduces data redundancy.

Timestamping solutions use counters to record the insertion of an element, instead of decrementing them periodically over time. The \emph{Detached Counting Bloom Filter Array}~\cite{dcba} associates a timer array to each of its filters, so to keep track of when data is inserted as well as when it needs to be retired. This scheme works well with the sliding window model, however, it is expensive in terms of memory use.

\subsection{Dictionary-based}

Dictionary-based approaches are mostly comparable to hash tables, but instead of storing the entire data, only a fingerprint of the element is saved. Although there are more BF-based schemes for duplicate detection, the ones with better performance results are dictionary-based, the most common ones being the \emph{Cuckoo Filter}~\cite{cf}, \emph{Morton Filter}~\cite{mf} and \emph{SWAMP}~\cite{swamp}.

The Cuckoo Filter is based on the Cuckoo Hash Table~\cite{cht} and consists of an array of buckets, where each can have multiple entries, and one entry is able to hold one fingerprint. An element has always two possible buckets to be stored in, determined by hash functions $h_1$ and $h_2$. To check for its presence, only the two candidate buckets for the item need to be queried.

The Morton Filter is somewhat similar to the previous method, but bias decisions in favor of $h_1$ instead and employs a compression strategy called the \emph{Block Store}. It improves, in terms of space usage and throughput, comparatively to the Cuckoo Filter.

SWAMP is the most recent state of the art dictionary-based approach. It functions as a cyclic buffer and maintains a TinyTable~\cite{tinytable} to keep track of the various fingerprints' frequencies. This scheme keeps the most recent data of the stream, by evicting the oldest entry when a new one is added, and is able to check, in constant time, if an element is present and how many distinct items are stored in the buffer.

\section{Age-Partitioned Bloom Filter}
\label{sec:apbf}

Many of the methods previously discussed either aim to optimize space utilization at the expense of using more complex algorithms, or are simple yet inefficient in terms of memory usage. To balance these properties (time complexity, space efficiency and algorithm complexity), \emph{Age-Partitioned Bloom Filters}~\cite{apbf} offer a BF-based data structure that improves over prior BF-based schemes and is able to compete with dictionary-based techniques.


\subsection{Structure}

An APBF follows the segmented approach and partitions the filter in a series of $k + l$ slices ($s_0, s_1, \ldots, s_{k+l-1}$), each with \emph{m} bits. This scheme also makes use of $k + l$ independent hash functions, one fixed per bit array, and maintains a counter \emph{n}, to keep track of how many elements have been inserted since the filter creation. Each desired false positive rate can be obtained by different combinations of $k$ and $l$, each combination providing a different trade-off in terms of operation speed and memory footprint. 

Like a Bloom Filter, an APBF has two basic operations: \emph{insert} and \emph{query}. However, unlike the latter, it can hold a specified window of the most recent elements and is able to expire stale information by shifting its slices.

\subsection{Insert}

Incoming data is stored in the first \emph{k} slices of the filter, by setting the corresponding \emph{k} bits to $1$. After some time, an insertion may trigger a shift and the slices move up in the array, i.e., $s_0$ becomes $s_1$, $s_1$ becomes $s_2$, and so on. Consequently, the last slice is discarded, to evict older elements, and a new one is added at location $0$, to receive more recent ones. In practice, slice $s_{k+l-1}$ is emptied out and is then reused as the new slice $s_0$.

The number of insertions made in the filter until a shift occurs is called a generation (\emph{g}). Slices shift whenever one of the \emph{k} slices reaches its maximum capacity. Note that, a slice hits an optimal use when its fill ratio (defined below) is equal to $1/2$.

\subsection{Query}
\label{subsec:apbf_query}

For an element to be deemed as present, it needs to be found in \emph{k} consecutive slices. The query algorithm starts at slice \emph{l} and moves up if a match is found while adding $1$ to the counter \emph{c} of consecutive matches. Otherwise, it goes down \emph{k} slices, saves the number of matches already found in the counter \emph{p}, resets \emph{c} and repeats the process once again. The algorithm terminates when the \emph{k} consecutive matches are found or when there are no more slices to check.

In this scheme, all elements inside the APBF sliding window are always guaranteed to be reported as present, i.e., no false negatives are observed.

\subsection{Fill ratio}
\label{subsec:apbf_fill_ratio}

The fill ratio \emph{r} of a slice is defined as the ratio of its set bits and depends on its size (\emph{m}) and on the number of elements it has stored (\emph{n}). Given this, it can be obtained by
\begin{equation}
\label{eq_fr}
    r = 1 - \left(1 - \frac{1}{m}\right)^n \approx 1 - e^{-\frac{n}{m}} \,.
\end{equation}

Due to shifting, slices have different fill ratios. The filter reaches a steady state after $k + l$ shifts, point when it stores between \emph{l} and $l + 1$ generations.
In the worst case, just before a shift occurs, the expected
fill ratios $r_0, r_1, \ldots, r_{k+l-1}$ are approximately given by
\begin{equation}
\label{eq_efr}
  r_i \approx
  \begin{cases}
    1 - 2^{-\frac{(i+1)}{k}} & \textrm{if } i < k, \\
    1/2 & \textrm{otherwise.}
  \end{cases}
\end{equation}

\section{Time-based Age-Partitioned Bloom Filter}
\label{sec:time_based_apbf}

The \emph{Time-based Age-Partitioned Bloom Filter} adapts the APBF model to hold a specified time window of elements. It is structured as a series of $k + l$ slices, each with $m_i$ bits and a fixed hash function. Only \emph{k} independent hash functions are used, seeing that slices that are apart by \emph{k} positions are not used for the same insertions and, therefore, can share the same hash function.

One of the things to take notice of in this new solution is the insertion rate. In the majority of cases (if not all), this parameter will not be known \textit{a priori}, so the filter must be able to adapt dynamically to its variation. The strategy is then to allow the filter to scale its number of slices, up and down, to adapt to the rate of insertions and be able to keep a time span of elements, while also adjusting the new slice $s_0$ size.

For a better understanding of the notations used in the following sections, Table~\ref{tab:notations} presents a list of variables and their meaning.

\begin{table}[t]
\centering
\caption{Notations.}
\label{tab:notations}
\resizebox{\linewidth}{!}{%
\begin{tabular}{cl}
\hline
\multicolumn{1}{|c|}{$\ $\textbf{Variable}$\ $} & \multicolumn{1}{c|}{\textbf{Description}} \\ \hline
\vspace{-0.3cm}  \\ \hline
\multicolumn{1}{|c|}{$s_i$} & \multicolumn{1}{l|}{Slice at location \emph{i}} \\ \hline
\multicolumn{1}{|c|}{$m_i$} & \multicolumn{1}{l|}{Size of slice \emph{i} in bits} \\ \hline
\multicolumn{1}{|c|}{$c_i$} & \multicolumn{1}{l|}{Number of elements slice \emph{i} can store (capacity)} \\ \hline
\multicolumn{1}{|c|}{$n_i$} & \multicolumn{1}{l|}{Number of elements inserted into slice \emph{i}} \\ \hline
\multicolumn{1}{|c|}{\emph{g}} & \multicolumn{1}{l|}{\begin{tabular}[c]{@{}l@{}}Number of elements inserted until a shift occurs\\ (generation size)\end{tabular}} \\ \hline
\multicolumn{1}{|c|}{$t_{span}$} & \multicolumn{1}{l|}{Time span specified by the user} \\ \hline
\multicolumn{1}{|c|}{$t_i$} & \multicolumn{1}{l|}{Timestamp of the last update made to slice \emph{i}} \\ \hline
\multicolumn{1}{|c|}{$u_i$} & \multicolumn{1}{l|}{\begin{tabular}[c]{@{}l@{}}Number of updates that can still be done until\\ slice \emph{i} gets full\end{tabular}} \\ \hline
\end{tabular}
}
\end{table}

\subsection{Slice life-time}

As previously stated, the time-based APBF should be able to add more slices, if needed, to accommodate the elements of a given time period. However, it must also be capable of retiring those same slices whenever they become stale. 
For a slice to be expired, its timestamp must be older than the time span. Therefore, slice $s_i$ is retired when
\begin{equation}
\label{eq_retire_slice}
    t_i < now() - t_{span} \,,
\end{equation}
where $now()$ is the timestamp in the present moment.
To avoid the cost of doing this at every insertion, checking whether a slice should be retired can be done only when a shift is triggered.

\subsection{Shift triggering}

The shifting mechanism guarantees that none of the first \emph{k} slices surpasses its maximum capacity. Now that slices can have different sizes, the number of updates a slice can take until it reaches its optimal use depends on its size, the number of insertions already received and the amount of shifts left until it gets to position \emph{k}, point when it stops receiving new elements. This means that the number of insertions left for slice \emph{i}, from that position, is distributed by the remaining $k - i$ shifts:
\begin{equation}
\label{eq_max_updates_slice}
    u_i = \lfloor \frac{m_i \times ln2 - n_i}{k - i} \rfloor \,.
\end{equation}
Therefore, the actual number of updates the filter can receive until a shift is triggered is
\begin{equation}
\label{eq_next_gen_size}
    g = \mathsf{min} \{ u_i \mid i \in [0, k-1]\}.
\end{equation}
This value is obtained right after a shift and can be saved on a counter that gets decremented by $1$ at every insertion, triggering a shift when it reaches $0$.

\subsection{Slice size}

After every shift, a new slice $s_0$ is added to the filter with size $m_0$. In the original APBF, this value is static and the same for all slices. However, in the time-based scheme, the size of $s_0$ can be updated according to the insertion rate.

The fraction $\frac{t_{span}}{l}$ represents the target time to have between shifts, so the number of slices remains constant ($= k + l$), $t_s$ is the time that passed since last shift and \emph{g} the number of insertions made in that period. The target generation size, i.e., the number of elements the filter should aim to insert between shifts, is then obtained by
\begin{equation}
\label{eq_target_gen_size}
    tg = \frac{g \times t_{span}}{t_s \times l} \,.
\end{equation}

The next step is to look at the remaining \emph{k} slices, from $1$ to $k - 1$, and calculate how many more updates can be done, at most, until (if ever) the new $s_0$ becomes the new limit. Applying Equation \ref{eq_next_gen_size} to slices $1$ to $k - 1$ gives the minimum number of updates possible until one of those slices reaches its full capacity. Consider $s_j$ to be the slice with the minimum amount of possible insertions. The new capacity for $s_0$ is given by
\begin{equation}
\label{eq_new_capacity}
    c_0 = count + i \times tg \, ,
\end{equation}
where \emph{count} is the total number of updates possible from position \emph{j} to \emph{k} and \emph{i} the number of shifts needed until $s_0$ reaches location \emph{j}. Therefore, the size of the new slice is given by
\begin{equation}
\label{eq_new_size}
    m_0 = \lceil \frac{c_0}{ln2} \rceil \,.
\end{equation}

\subsection{Query}

The query algorithm of this new scheme follows the same logic as the original APBF, detailed in Section~\ref{subsec:apbf_query}, the only difference being an additional check to see if the slices, that are being queried for the element, are still within the specified time period. If not, then they are not considered in the search. Algorithm~\ref{lst:time_apbf_query} shows the detailed process of the query operation.

\begin{lstlisting}[float=t, language=C, caption={Query algorithm.}, label={lst:time_apbf_query}, escapeinside=||, captionpos=b]
|$\textbf{function}$ query(x)|
    |$i := numSlices - k$, $p := 0$, $c := 0$|
    |$\textbf{while}$ $i \geq 0$ $\textbf{do}$| 
        |$\textbf{if}$ $s_i[h_i(x)] = 1 \textbf{ and } t_i \geq now() - t_{span}$ $\textbf{then}$|
            |$c := c + 1$, $i := i + 1$|
            |$\textbf{if}$ $p + c = k$ $\textbf{then}$|
                |$\textbf{return}$| true
        |$\textbf{else}$|
            |$i := i - k$, $p := c$, $c := 0$|
    |$\textbf{return}$| false
\end{lstlisting}

\section{Experimental Evaluation}
\label{sec:case_studies}

In this section, the time-based APBF is evaluated in terms of the number of slices in the filter, the size of $s_0$, the memory use per element of interest and the false positive rate. For this purpose, a C implementation of this scheme was developed and is publicly available 
at \href{https://github.com/RedisBloom/RedisBloom/tree/AgePartitionedBF}{Redis Bloom} in the \emph{AgePartitionedBF} branch.

\subsection{Number of slices and size of slice 0}

As previously stated, the number of slices in a time-based APBF can increase to accommodate more data when necessary, but should also decrease back to the base value of $k + l$ when the filter stabilizes. Regarding the size of $s_0$, it should adapt to the insertion rate, by increasing and decreasing when needed, and remain constant when the filter reaches a steady state.

\begin{figure}[tbp]%
    \centering
    \subfloat[]{\includegraphics[width=\linewidth]{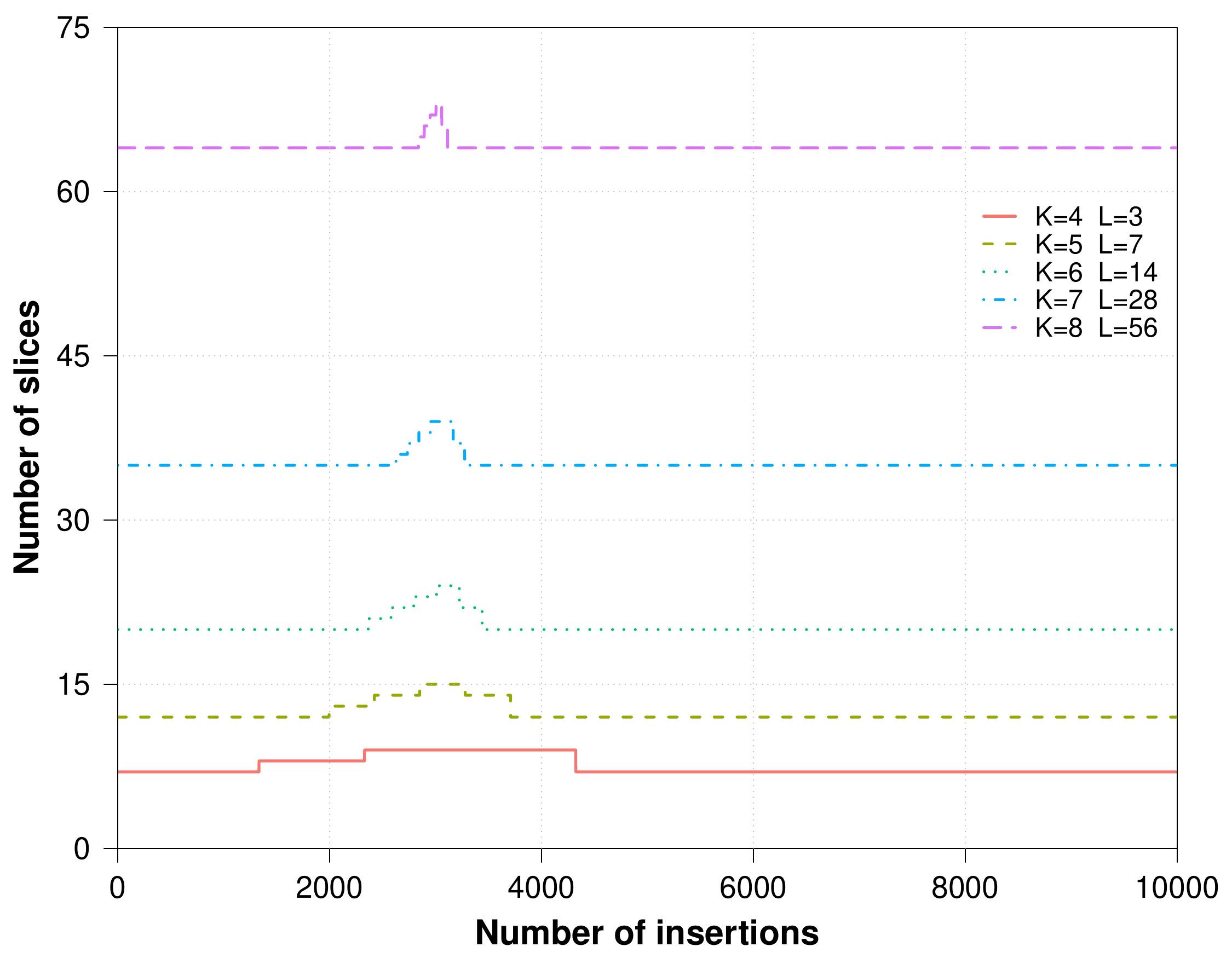}}
    \hspace{0.5cm}
    \subfloat[]{\includegraphics[width=\linewidth]{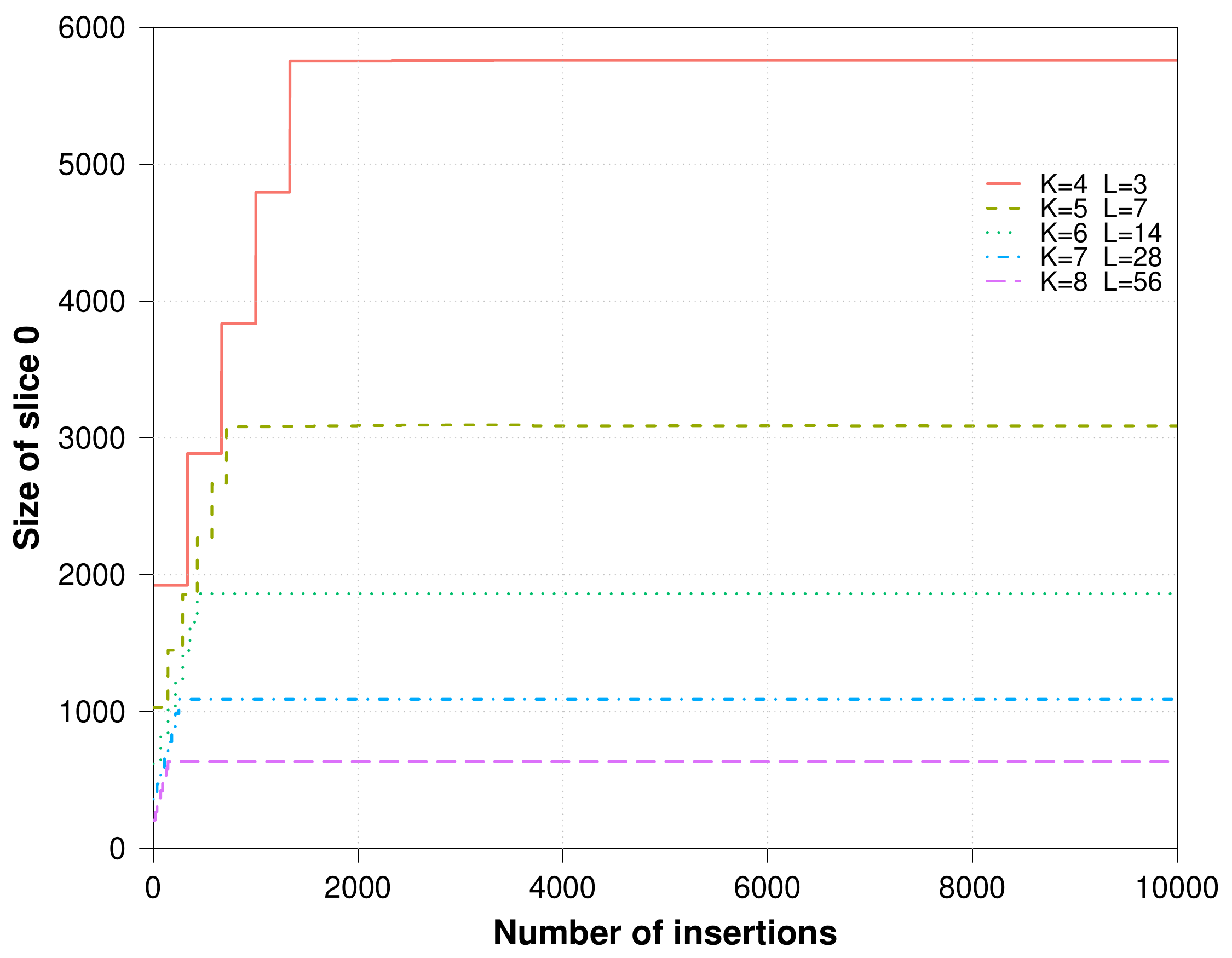}}
    \caption{(a) Number of slices, at every insertion, of each filter with an initial capacity of $1000$; (b) Size of slice $0$, at every insertion, of each filter with an initial capacity of $1000$.}
    \label{fig:slices_size_under}%
\end{figure}

\begin{figure}[tbp]%
    \centering
    \subfloat[]{\includegraphics[width=\linewidth]{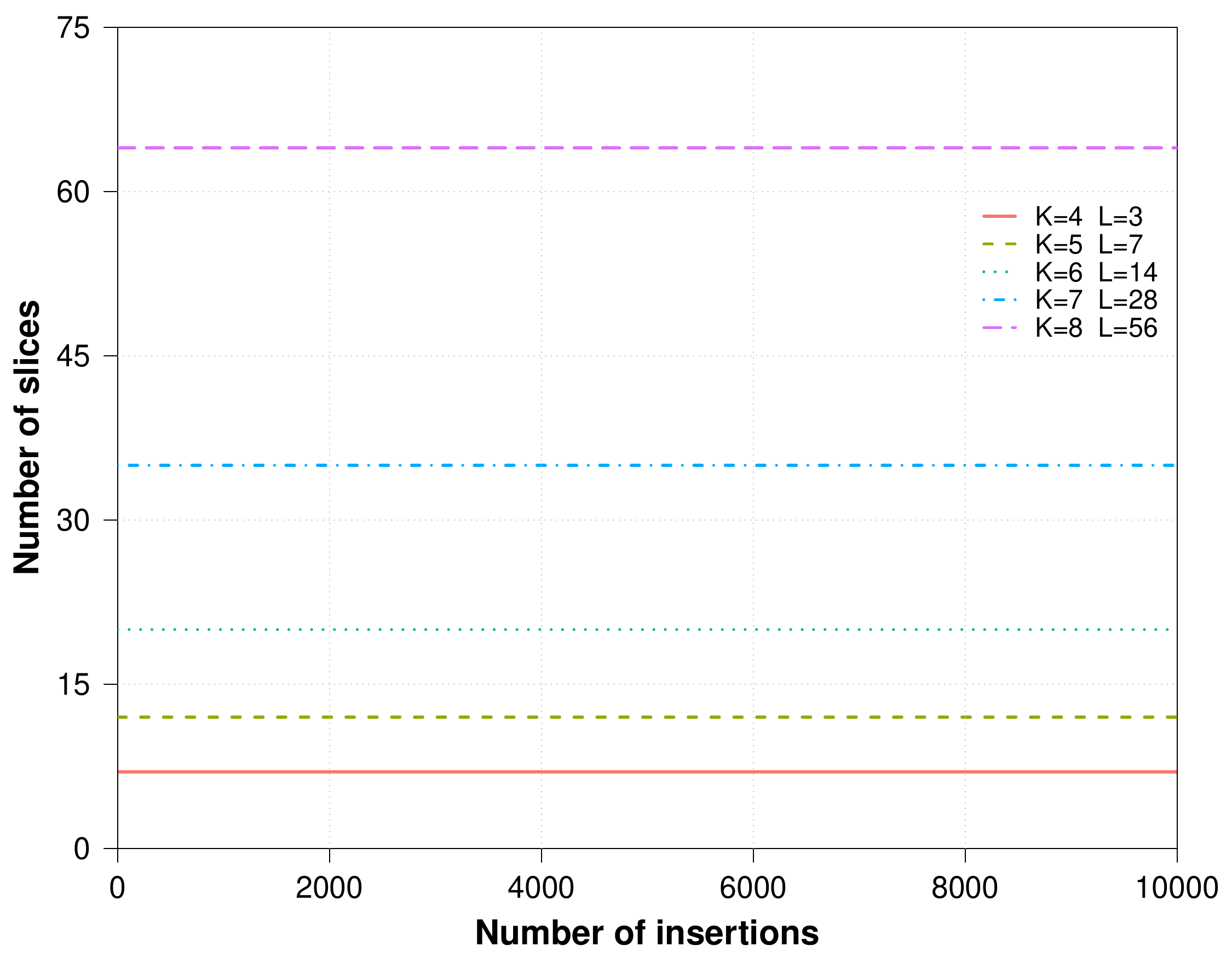}}%
    \hspace{0.5cm} 
    \subfloat[]{\includegraphics[width=\linewidth]{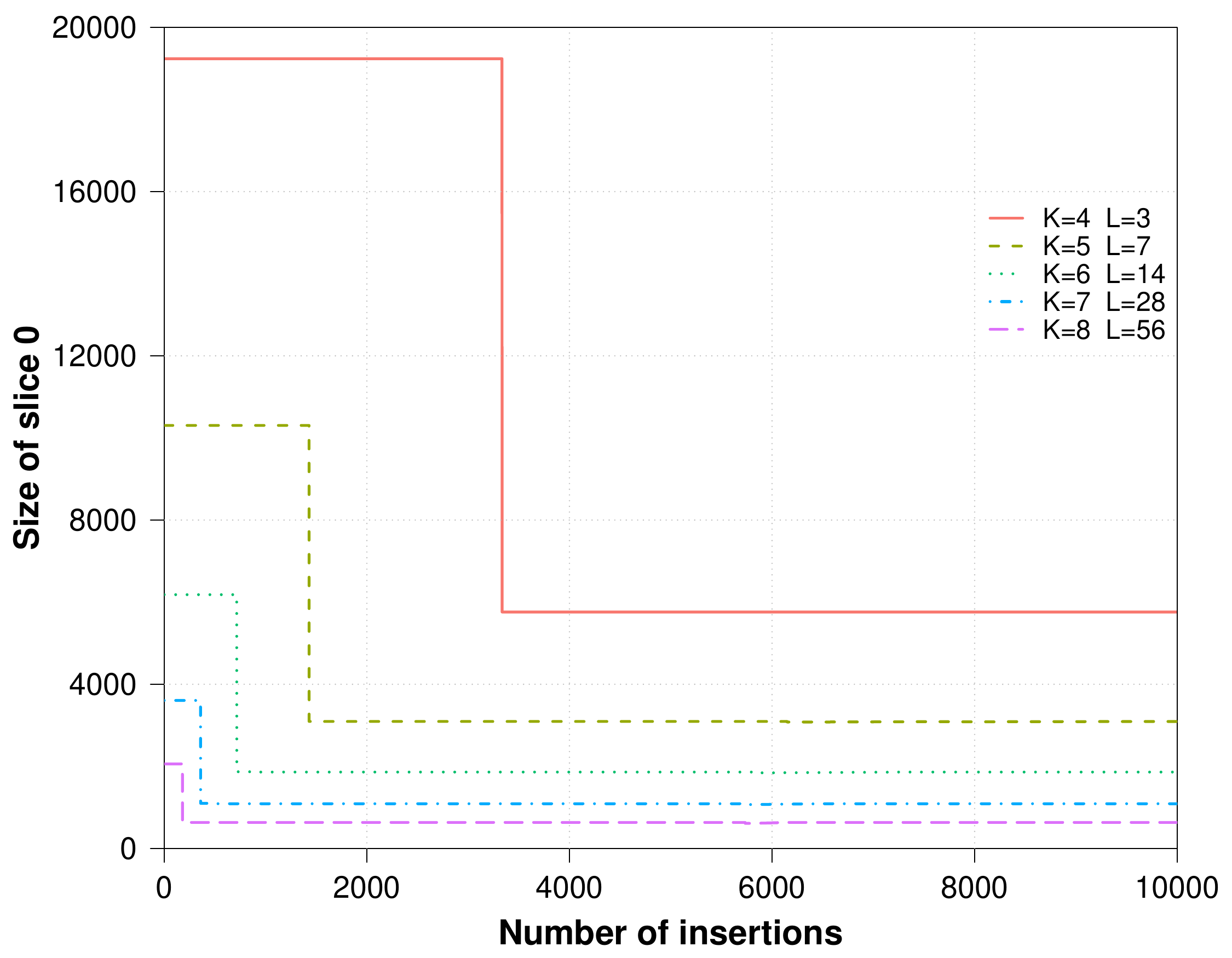}}%
    \caption{(a) Number of slices, at every insertion, of each filter with an initial capacity of $10000$; (b) Size of slice $0$, at every insertion, of each filter with an initial capacity of $10000$.}
    \label{fig:slices_size_over}%
\end{figure}

In this experiment, a total of $10$ filters, with the same error rate of $0.1$, a time span of $300$ seconds (equivalent to $5$ minutes), an insertion rate of $0.1$ seconds, and different initial capacities and \emph{k} and \emph{l} combinations, were subject to a data stream of $10000$ distinct elements. Figures \ref{fig:slices_size_under} and \ref{fig:slices_size_over} show the results of measuring the number of slices and the size of slice $0$, at every insertion, when the filters are under and over-dimensioned, respectively.

In the first case, when the initial capacity is $1000$, the filters are under-dimensioned and so an increase of the number of slices, as well as the size of slice $0$, is observed, as expected. When the filters stabilize, i.e., start inserting the exact aimed number of elements between shifts, the number of slices begins to decrease back to the base value of $k + l$ and the size of $s_0$ remains constant after reaching its optimal value.

When the filters are over-dimensioned, the number of slices doesn't suffer any alterations, since slices have more space allocated than necessary and, therefore, there's no need to add more. However, at the very first shift, the size of $s_0$ is updated down to the optimal value to conserve the memory space and avoid unnecessary waste.

Furthermore, by looking at the filters with ($k = 4$, $l = 3$) and ($k = 8$, $l = 56$), it's safe to say that filters with lower \emph{k} and \emph{l} combinations take longer to stabilize, since their slices are larger and, thus, take longer to fill and, consequently, to shift, which is the point when the size of slice $0$ is updated.

\subsection{Memory use per element}

\begin{figure}[tb]%
    \centering
    \subfloat[]{\includegraphics[width=\linewidth]{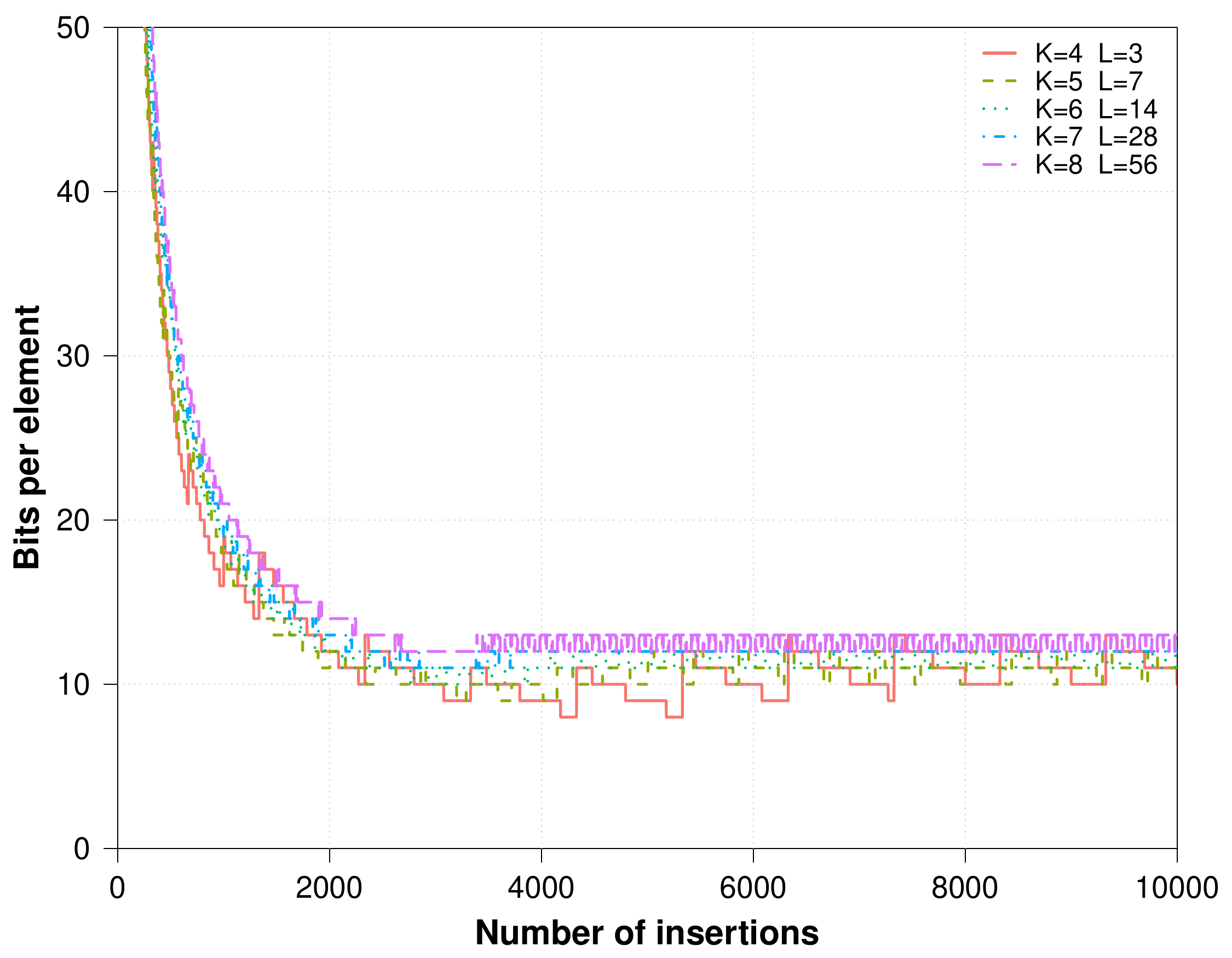}}%
    \hspace{0.5cm}
    \subfloat[]{\includegraphics[width=\linewidth]{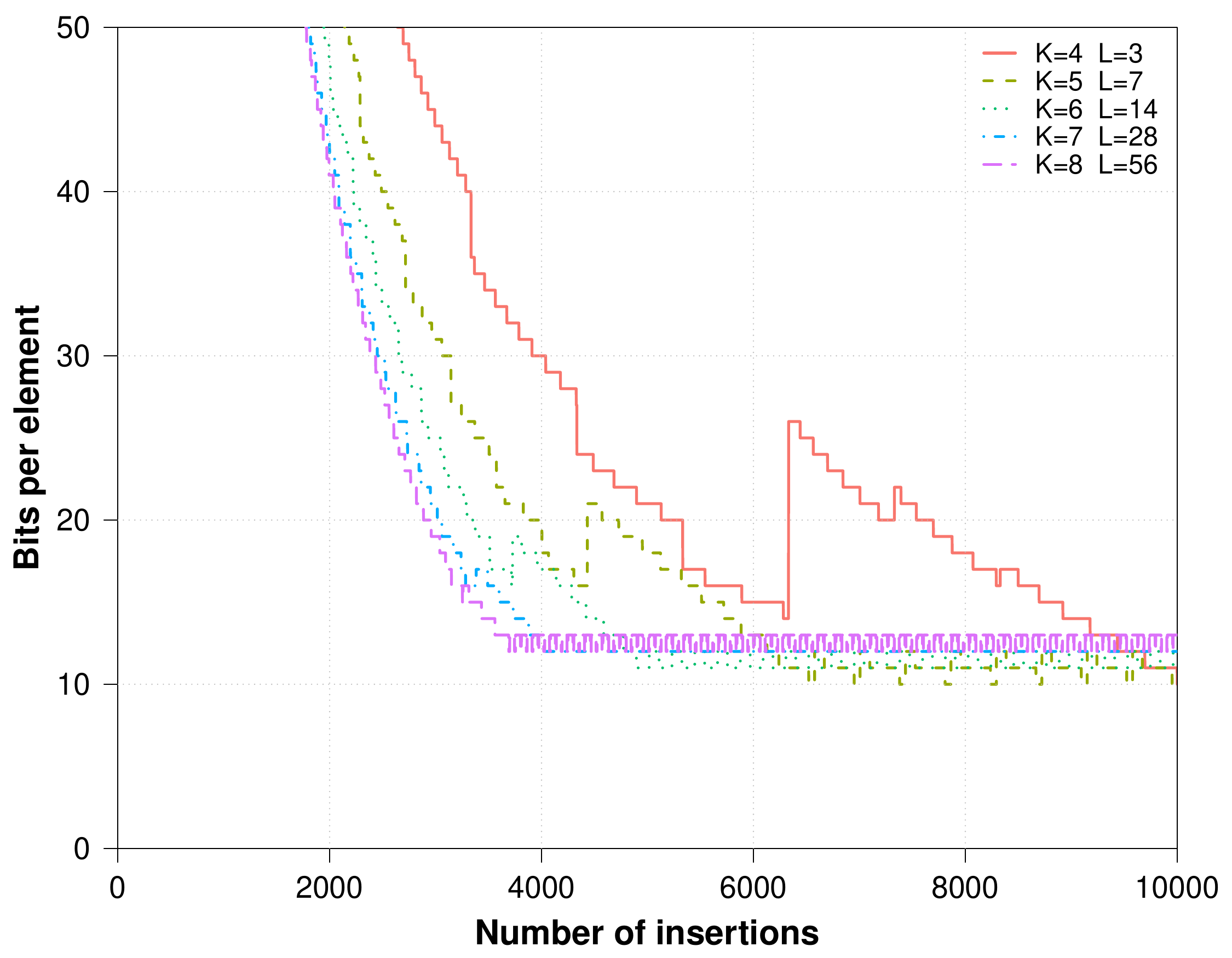}}%
    \caption{(a) Bits per element, at every insertion, of each filter with an initial capacity of $1000$; (b) Bits per element, at every insertion, of each filter with an initial capacity of $10000$.}
    \label{fig:bits_per_element}%
\end{figure}

Another important metric to analyze is how many bits are allocated to each stored element inside the target time window. In this experiment, a total of $10$ filters, with the same error rate of $0.1$, a time span of $300$ seconds (equivalent to $5$ minutes), an insertion rate of $0.1$ seconds, and different initial capacities and \emph{k} and \emph{l} combinations, were subject to a data stream of $10000$ distinct elements. Fig. \ref{fig:bits_per_element} shows the results of measuring the bits per element, at every insertion, when the filters are under and over-dimensioned, respectively.

As expected, over-dimensioned filters take longer to reach a steady state due to their initial over-sized slices. Still, in both cases, the number of bits per element stabilizes around values between $10$ and $13$.

When analyzing this metric for different error rates (Table \ref{tab:bits_errors}), an increase in the range of bits used per element is observed for higher precisions.

\begin{table}[tb]
\centering
\caption{Range of bits per element allocated for each error rate.}
\label{tab:bits_errors}
\resizebox{\linewidth}{!}{%
\begin{tabular}{|c|c|c|c|c|c|c|}
\hline
\multicolumn{2}{|c|}{\textbf{Error rate}} & $\ $0.1$\ $ & $\ $0.01$\ $ & $\ $0.001$\ $ & $\ $0.0001$\ $ & $\ $0.00001$\ $ \\ \hline
\multirow{2}{*}{\textbf{\begin{tabular}[c]{@{}c@{}}$\ $Bits per$\ $\\ element\end{tabular}}} & $\ $\textbf{Min}$\ $ & 10 & 19 & 26 & 32 & 41 \\ \cline{2-7} 
 & \textbf{Max} & 13 & 24 & 35 & 45 & 56 \\ \hline
\end{tabular}
}
\end{table}

\subsection{False positive rate}

\begin{figure}[tbp]%
    \centering
    \subfloat[]{\includegraphics[width=\linewidth]{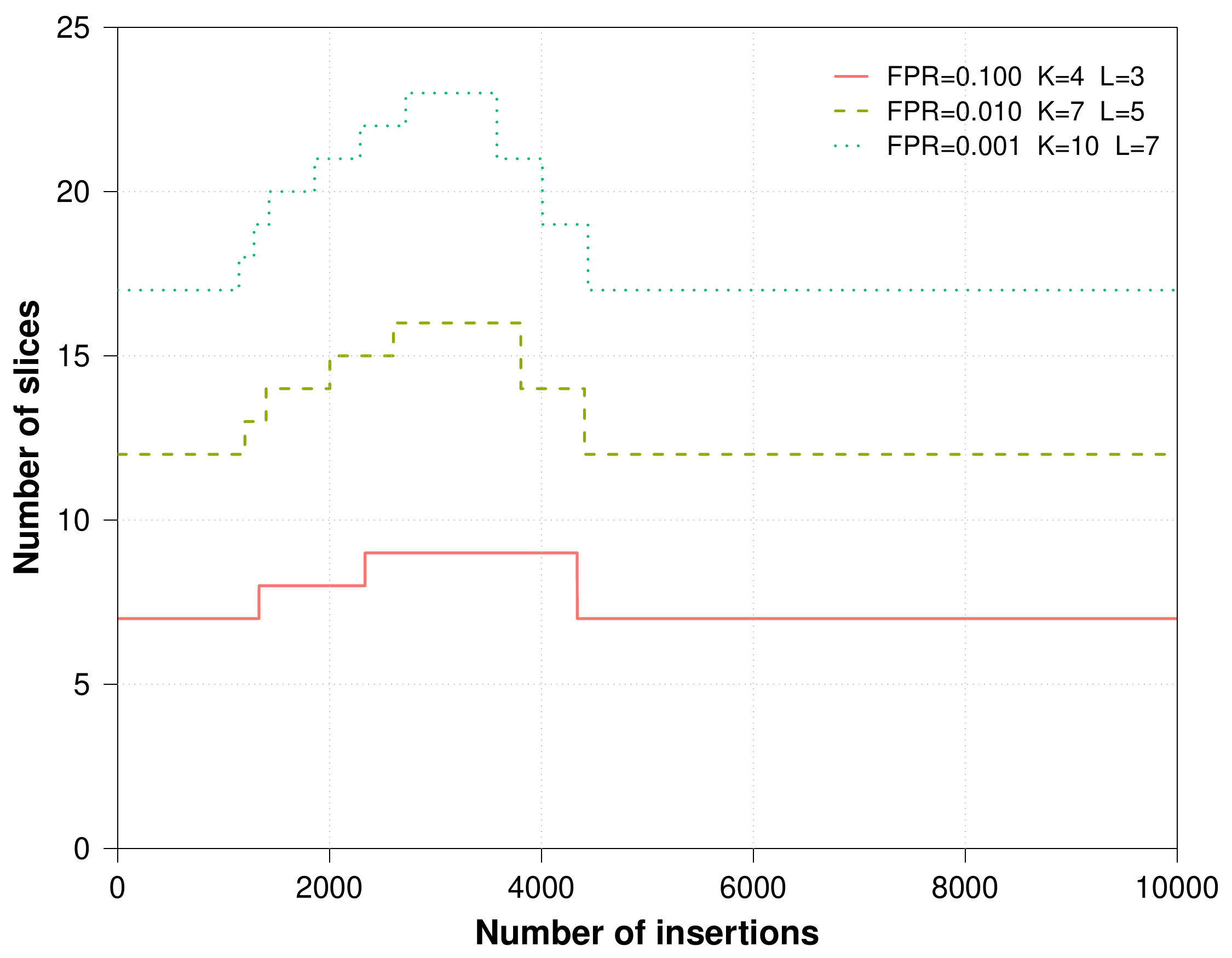}}%
    \hspace{0.5cm}
    \subfloat[]{\includegraphics[width=\linewidth]{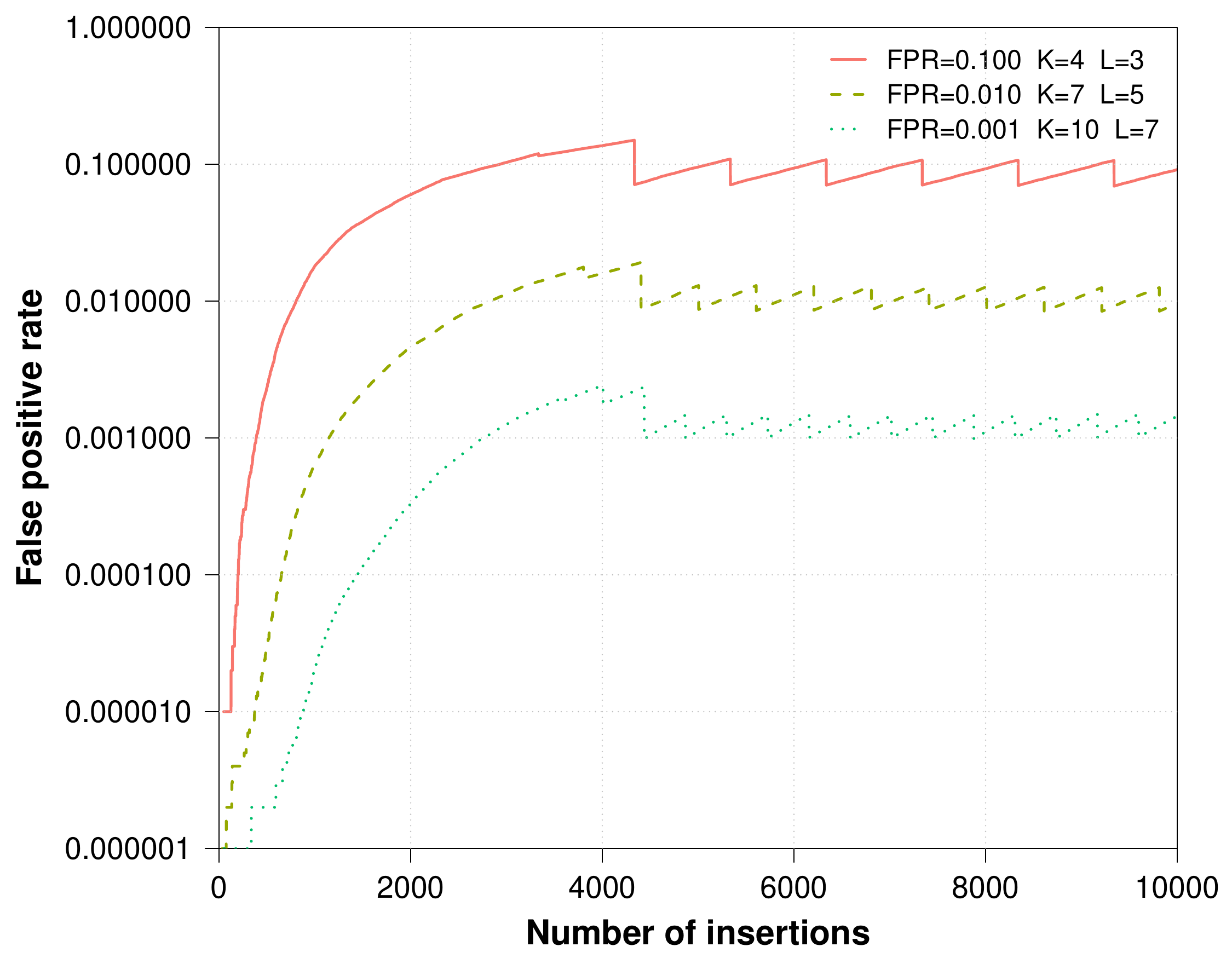}}%
    \caption{(a) Number of slices, at every insertion, of each filter with an initial capacity of $1000$; (b) False positive rate, at every insertion, of each filter with an initial capacity of $1000$.}
    \label{fig:fpr_under}%
\end{figure}

\begin{figure}[tbp]%
    \centering
    \subfloat[]{\includegraphics[width=\linewidth]{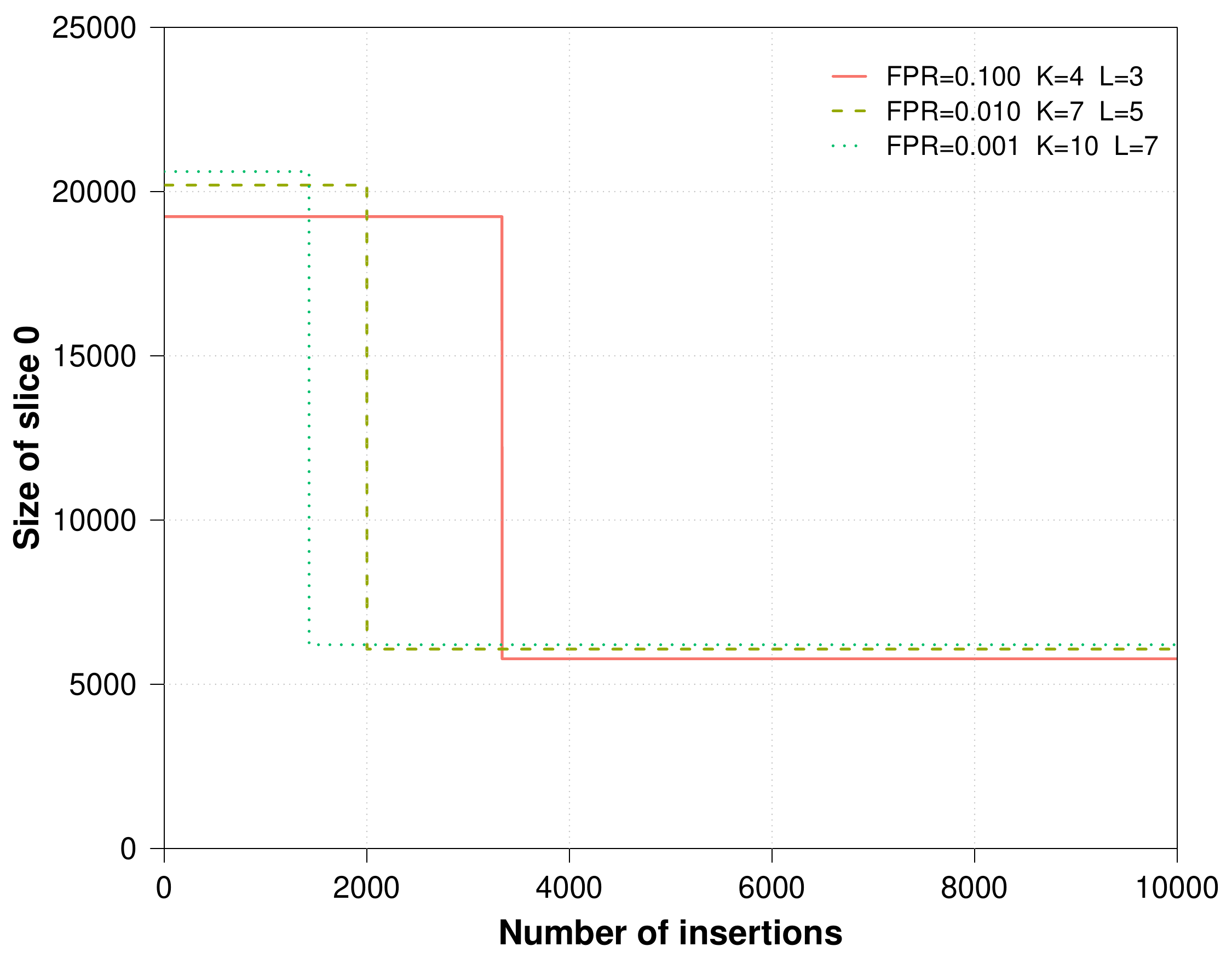}}%
    \hspace{0.5cm}
    \subfloat[]{\includegraphics[width=\linewidth]{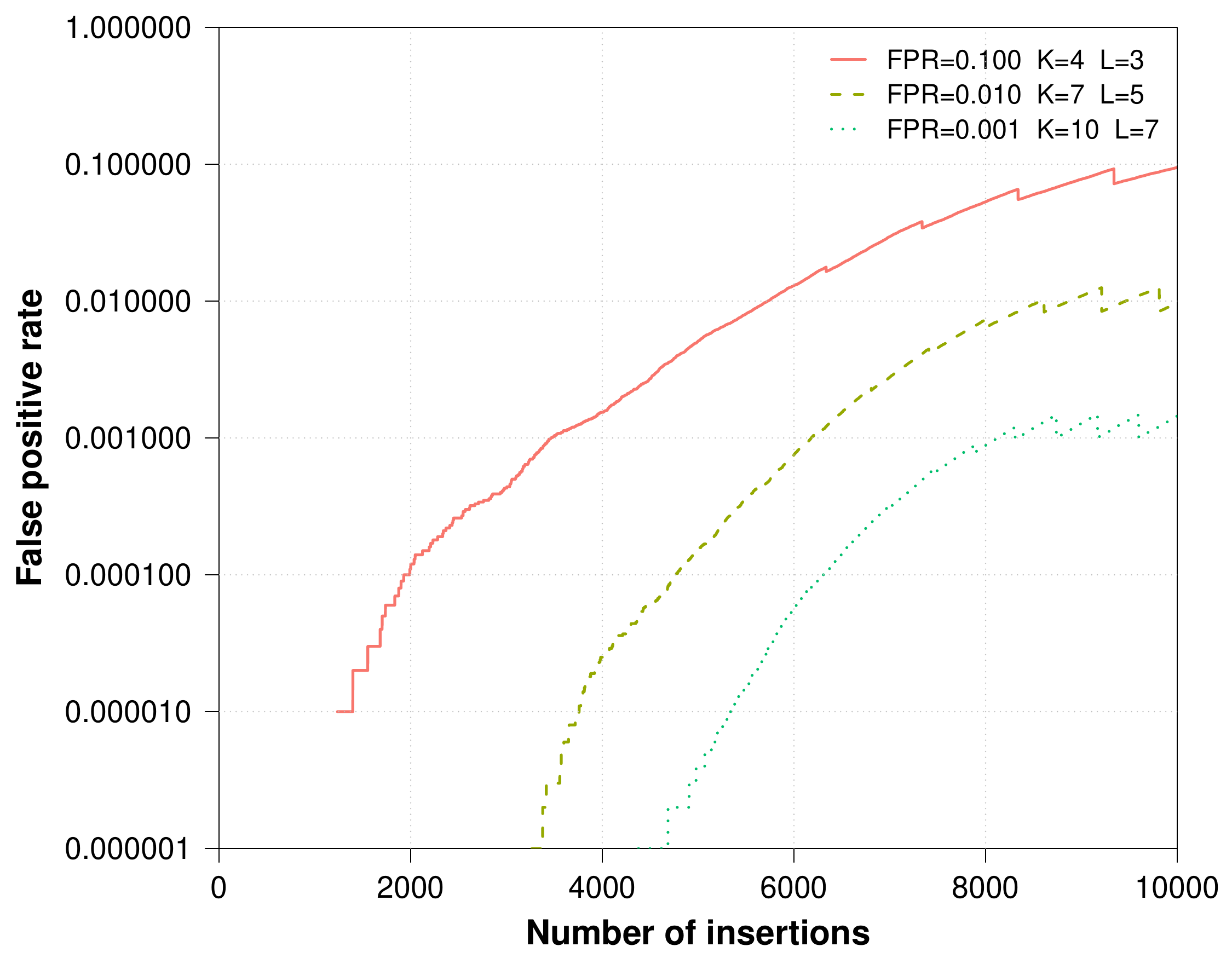}}%
    \caption{(a) Size of slice $0$, at every insertion, of each filter with an initial capacity of $10000$; (b) False positive rate, at every insertion, of each filter with an initial capacity of $10000$.}
    \label{fig:fpr_over}%
\end{figure}

To measure the false positive rate (FPR), a total of $6$ filters, with the same time span of $300$ seconds (equivalent to $5$ minutes), an insertion rate of $0.1$ seconds, and different error rates, initial capacities and \emph{k} and \emph{l} combinations, were subject to a data stream of $10000$ distinct elements. At every insertion, each filter with an error rate of $1/10^i$ was queried for $10^i \times 10000$ distinct elements, known not to be present. Figures \ref{fig:fpr_under} and \ref{fig:fpr_over} show the results of measuring the FPR, at every insertion, when the filters are under and over-dimensioned, respectively.

As seen before, when filters are under-dimensioned, the number of slices increases to make room for more information. From Fig. \ref{fig:fpr_under}, it's possible to affirm that this increase in the amount of slices doesn't affect much the FPR, with it just going a bit over the predefined threshold.

When an over-dimensioning of the filter happens, the false positive rate takes longer to stabilize and reach the configured maximum rate because of the initial over-sized slices. Since they're too big, they'll never reach their target fill ratio and, therefore, the probability of finding a false positive is lower. Only after a number of shifts, when those initial slices are discarded, does the FPR reach the predefined threshold.

In both cases, as the data is inserted, the FPR stabilizes around the configured error rate. The zig zag effect seen is the result of filling up the slices (the peak represents the state right before a shift) and of discarding the last one and adding a new empty slice at position $0$ (the lower end represents the state just after a shift). This effect is attenuated in higher \emph{k} and \emph{l} combinations, as well as in higher precisions.

A simple test was also made to confirm that all elements inserted within the specified time span were correctly identified as present and, as intended, no false negatives were registered.

\section{Conclusion}
\label{sec:conclusion}

In this paper, we presented the \emph{Time-limited Bloom Filter}, a segmented-based approach that partitions the filter in $k + l$ slices. When necessary, this data structure can increase its number of slices, so as to accommodate more data, and adapt the size of slice $0$ accordingly. Symmetrically, slices can also be retired when their data becomes stale, i.e., when it no longer belongs to the specified time span. Furthermore, slices that are apart by \emph{k} positions can share the same hash function, since they will not be used for the same insertions, and so, only \emph{k} hash functions need to be used for this scheme. 
Elements that are inserted within the time window are always reported as present, which means this solution has no false negatives. Also, the false positive rate stabilizes around the predefined maximum rate. Even when the number of slices increases, the FPR doesn't jump abruptly, only goes slightly above the configured error rate.

Another interesting metric analyzed was the memory use per element belonging to the target time window. It was observed that lower precisions use fewer bits per item, between $10$ and $13$, and that as the error rate decreases the memory use increases, reaching values between $41$ and $56$ bits per element in the highest precision.

Regarding the slices retirement, the minimum value the number of slices of a time-based APBF can decrease to, in this work, is $k + l$. However, potentially it is possible to decrease the number of slices as low as \emph{k} without affecting the false positive rate, an interesting aspect to analyse in the future.

The mechanism presented in this paper was implemented in C and is available as a Redis module, loadable into a Redis server instance, and can be used with the command line Redis client or from client libraries in several languages.  Implementation is available at 
\url{https://github.com/RedisBloom/RedisBloom/tree/AgePartitionedBF}.


\bibliographystyle{plain}
\bibliography{paper}

\end{document}